\documentclass{article}%
\usepackage{graphicx}
\usepackage{amsmath}%
\usepackage{amsfonts}%
\usepackage{amssymb}
\newtheorem{theorem}{Theorem}
\newtheorem{acknowledgement}[theorem]{Acknowledgement}

\begin{document}

\title{{\Large Non-Volkov solutions for a charge in a plane wave}}
\author{V. Bagrov\thanks{On leave from Tomsk State University and Tomsk Institute of
High Current Electronics, Russia; e-mail: bagrov@phys.tsu.ru}\thinspace
\thinspace\ and D. Gitman\thanks{E-mail: gitman@dfn.if.usp.br}\\ \\Instituto de F\'{\i}sica, Universidade de S\~{a}o Paulo,\\Caixa Postal 66318-CEP, 05315-970 S\~{a}o Paulo, S.P., Brazil}
\maketitle

\begin{abstract}
We focus our attention, once again, on the Klein--Gordon and Dirac equations
with a plane-wave field. We recall that for the first time a set of solutions
of these equations was found by Volkov. The Volkov solutions are widely used
in calculations of quantum effects with electrons and other elementary
particles in laser beams. We demonstrate that one can construct sets of
solutions which differ from the Volkov solutions and which may be useful in
physical applications. For this purpose, we show that the transversal charge
motion in a plane wave can be mapped by a special transformation to
transversal free particle motion. This allows us to find new sets of solutions
where the transversal motion is characterized by quantum numbers different
from Volkov's (in the Volkov solutions this motion is characterized by the
transversal momentum). In particular, we construct solutions with
semiclassical transversal charge motion (transversal squeezed coherent
states). In addition, we demonstrate how the plane-wave field can be
eliminated from the transversal charge motion in a more complicated case of
the so-called combined electromagnetic field (a combination of a plane-wave
field and constant colinear electric and magnetic fields). Thus, we find new
sets of solutions of the Klein--Gordon and Dirac equations with the combined
electromagnetic field.

\end{abstract}

\section{Introduction}

Relativistic wave equations (Dirac and Klein--Gordon) provide a basis for
relativistic quantum mechanics and QED of spinor and scalar particles. In
relativistic quantum mechanics, solutions of relativistic wave equations are
referred to as one-particle wave functions of fermions and bosons in external
electromagnetic fields. In QED, such solutions permit the development of the
perturbation expansion known as the Furry picture, which incorporates the
interaction with the external field exactly, while treating the interaction
with the quantized electromagnetic field perturbatively
\cite{Schwe61,GreMuR85,BagGi90,FraGiS91,Grein97}. The most important exact
solutions of the Klein--Gordon and Dirac equations are: solutions with the
Coulomb field, which allow one to construct the relativistic theory of atomic
spectra \cite{BetSa57}, solutions with a uniform magnetic field, which provide
the basis of synchrotron radiation theory \cite{synchrotron}, and solutions in
the field of a plane wave, which are widely used for calculations of quantum
effects involving electrons and other elementary particles in laser beams
\cite{laser}. This is why any progress in studying these basic solutions can
result in new physical applications and seems to be important. In the present
article, we focus our attention, once again, on the Klein--Gordon and Dirac
equations with a plane-wave field. We recall that for the first time a set of
solutions of these equations was obtained by Volkov in \cite{Volkov}, see Sec.
2. It is Volkov's solutions that have been used in all of the above-mentioned
calculations. However, we demonstrate below that one can construct sets of
solutions which differ from the Volkov solutions, and which may be useful in
physical applications. It is known that the transversal charge motion in the
Volkov solutions is characterized by a definite transversal momentum. We show
that the transversal charge motion in a plane wave can be mapped by a special
transformation to transversal free particle motion. This allows us to find
sets of solutions where the transversal charge motion is characterized by
different (from the Volkov case) quantum numbers, see Sec. 3. The importance
of constructing solutions with different quantum numbers is related to
possible different experimental conditions, e.g., special initial charge
states, or specially prepared charge states in a plane-wave. In particular, we
construct solutions with semiclassical transversal charge motion (transversal
coherent states). In Sec. 4, we demonstrate how the plane-wave field can be
eliminated from the transversal charge motion in a more complicated case of
the so-called combined electromagnetic field (a combination of a plane-wave
field and colinear constant electric and magnetic fields). One ought to say
that Volkov-like solutions with the combined field were first obtained in
\cite{12,13}. Using the above-mentioned transformation, we find new sets of
solutions of the Klein--Gordon and Dirac equations with the combined
electromagnetic field.

\section{Volkov solutions for a charge in a plane-wave}

An electromagnetic field of a plane-wave propagating along a unit vector
$\mathbf{n}$ (here and elsewhere, we choose $\mathbf{n}=\left(  0,0,1\right)
$) can be described by potentials $A^{\mu}=A^{\mu}\left(  \xi\right)  ,$ where
$\xi=x^{0}-x^{3}=nx\,$is a light-cone variable\footnote{As usual, we denote
the Minkowski coordinates by $x=\left(  x^{\mu}\right)  =\left(
x^{0},\mathbf{r}\right)  $,\ $\mathbf{r}=\left(  x^{i},\,i=1,2,3\right)  $.}
($n^{\mu}=\left(  1,0,0,1\right)  ,\;n^{2}=0)$. Choosing the Lorentz gauge
$\partial_{\mu}A^{\mu}=0,$ which implies that $nA=0,$ and the gauge condition
$A_{0}=0,$ we have $A^{\mu}=\left(  0,\mathbf{A}\right)  \,,\;\mathbf{nA}%
=0\,.$ The electric $\mathbf{E}$ and magnetic $\mathbf{H}$ fields are
expressed through the potentials as $\mathbf{E}=-\mathbf{A}^{\prime}%
$,\ $\mathbf{H}=\left[  \mathbf{A}^{\prime}\mathbf{\times n}\right]  .$ Here
and elsewhere, primes stand for derivatives with respect to $\xi,$ i.e.,
$\mathbf{A}^{\prime}=d\mathbf{A/}d\xi$ .

In the case under consideration, the Lorentz equations have the form
\begin{equation}
m\ddot{x}^{\mu}=e\left(  \dot{x}A^{\prime}\right)  n^{\mu}-e\dot{A}^{\mu
}\Longleftrightarrow\left\{
\begin{array}
[c]{l}%
m\ddot{x}^{0}=-e\left(  \mathbf{A}^{\prime}\,\mathbf{\dot{r}}\right)  \\
m\mathbf{\ddot{r}}=-e\left[  \mathbf{A}^{\prime}\dot{\xi}+\mathbf{n}\left(
\mathbf{A}^{\prime}\,\mathbf{\dot{r}}\right)  \right]
\end{array}
\right.  ,\label{pw.8}%
\end{equation}
where dots stand for derivatives with respect to the proper time $\tau$. After
multiplying these equations by the vector $n^{\mu}$, we obtain $m\ddot{\xi
}=0\Longrightarrow m\dot{\xi}=\lambda\,,$ where $\lambda$ is an integral of
motion. The classical action (a solution of the Hamilton--Jacobi equation)
that depends on the integrals of motion $\lambda$ and $\mathbf{p}_{\perp}$ has
the form \cite{LanLi78}%
\begin{equation}
S\left(  x\right)  =-\frac{\lambda}{2}\left(  x^{0}+x^{3}\right)
+\mathbf{p}_{\perp}\mathbf{r}-\frac{1}{2\lambda}\int\left(  m^{2}%
+\mathbf{\mathbf{P}_{\bot}^{2}}\right)  d\xi\,,\label{pw.20}%
\end{equation}
where%
\begin{align}
&  \lambda=np=nP\,,\;p_{\mu}=-\partial_{\mu}S\,,\;P_{\mu}=p_{\mu}-eA_{\mu
}\,,\;\mathbf{p=}\left(  -p_{1},-p_{2},-p_{3}\right)  \,,\nonumber\\
&  \mathbf{P}_{\bot}=\mathbf{p}_{\perp}-e\mathbf{A}\left(  \xi\right)
,\;\mathbf{p}_{\perp}=\left(  -p_{1},-p_{2},0\right)  \mathbf{,\;p}=\nabla
S=\mathbf{p}_{\perp}+\frac{m^{2}+\mathbf{\mathbf{P}_{\bot}^{2}}-\lambda^{2}%
}{2\lambda}\mathbf{n\,.}\label{pw.21}%
\end{align}

Considering the Klein--Gordon and Dirac equations with a plane-wave field,%
\begin{align}
\mathcal{\hat{K}}\varphi\left(  x\right)   &  =0\,,\;\mathcal{\hat{K}}=\hat
{P}^{2}-m^{2}\,,\;\hat{P}_{\mu}=\hat{p}_{\mu}-eA_{\mu}\,,\label{pw.22}\\
\mathcal{\hat{D}}\psi\left(  x\right)   &  =0\,,\;\mathcal{\hat{D}}=\hat
{P}_{\mu}\gamma^{\mu}-m\,,\label{pw.26}%
\end{align}
one usually seeks for such solutions that are eigenvectors of the operators
(quantum integrals of motion) $\hat{\lambda}$ and $\mathbf{\hat{p}}_{\perp},$
which commute both with $\mathcal{\hat{K}}$ and $\mathcal{\hat{D}}$ and
between themselves:%
\begin{align}
&  \hat{\lambda}=\left(  n\hat{p}\right)  \,,\mathbf{\;}\hat{p}_{\mu
}=i\partial_{\mu}\,\mathbf{,}\;\mathbf{\hat{p}}=-i\nabla,\;\mathbf{\hat{p}%
}_{\perp}=-i\left(  \partial_{1},\partial_{2},0\right)  \mathbf{\,}%
,\nonumber\\
&  \left[  \hat{\lambda},\mathbf{\hat{p}}_{\perp}\right]  =\left[
\hat{\lambda},\mathcal{\hat{K}}\right]  =\left[  \hat{\lambda},\mathcal{\hat
{D}}\right]  =\left[  \mathbf{\hat{p}}_{\perp},\mathcal{\hat{K}}\right]
=\left[  \mathbf{\hat{p}}_{\perp},\mathcal{\hat{D}}\right]  =0\,.\label{pw.23}%
\end{align}
Such solutions where first obtained by Volkov, see \cite{Volkov}, and are
referred to as the Volkov solutions in what follows. The Volkov solutions are
subject to the conditions%
\begin{align}
\hat{\lambda}\varphi_{\lambda,\mathbf{p}_{\perp}}\left(  x\right)   &
=\lambda\varphi_{\lambda,\mathbf{p}_{\perp}}\left(  x\right)
\,,\;\mathbf{\hat{p}}_{\perp}\varphi_{\lambda,\mathbf{p}_{\perp}}\left(
x\right)  =\mathbf{p}_{\perp}\varphi_{\lambda,\mathbf{p}_{\perp}}\left(
x\right)  \,,\nonumber\\
\hat{\lambda}\psi_{\lambda,\mathbf{p}_{\perp}}\left(  x\right)   &
=\lambda\psi_{\lambda,\mathbf{p}_{\perp}}\left(  x\right)  \,,\;\mathbf{\hat
{p}}_{\perp}\psi_{\lambda,\mathbf{p}_{\perp}}\left(  x\right)  =\mathbf{p}%
_{\perp}\psi_{\lambda,\mathbf{p}_{\perp}}\left(  x\right)  \,,\label{pw.24}%
\end{align}
without restrictions on $\lambda$, and have the form\footnote{We denote the
Pauli matrices as $\mathbf{\sigma}=\left(  \sigma_{i},\,i=1,2,3\right)  $.}%
\begin{align}
\varphi_{\lambda,\mathbf{p}_{\perp}}\left(  x\right)   &  =N\exp iS\left(
x\right)  \,,\label{pw.25}\\
\psi_{\lambda,\mathbf{p}_{\perp}}\left(  x\right)   &  =N\exp iS\left(
x\right)  \left(
\begin{array}
[c]{c}%
m+\lambda+\sigma_{3}\left(  \mathbf{\sigma}\,\mathbf{\mathbf{P}_{\bot}%
}\right)  \\
\left(  m-\lambda\right)  \sigma_{3}+\mathbf{\sigma}\,\mathbf{\mathbf{P}%
_{\bot}}%
\end{array}
\right)  \vartheta\,,\label{pw.35}%
\end{align}
where $S\left(  x\right)  $ is the classical action (\ref{pw.20}), $N$ is a
normalization constant, and $\vartheta$ is an arbitrary two-component constant spinor.

The set of solutions (\ref{pw.25}) and (\ref{pw.35}) is orthonormal, with
respect to the scalar products on the null-plane $\xi=\mathrm{const}$, and
with respect to the scalar products on the plane $x^{0}=\mathrm{const.}$

\section{Non-Volkov solutions}

The Volkov solutions are subject to the conditions (\ref{pw.24}), i.e., they
represent quantum states with the conserved integrals of motion $\lambda$ and
$\mathbf{p}_{\perp}.$ We present below a different way of solving the
Klein--Gordon and Dirac equations with a plane-wave field. In such a way, we
obtain a wider class of solutions. In particular, the latter do not have to be
eigenfunctions of the operators $\mathbf{\hat{p}}_{\perp}$.

\subsection{Exclusion of plane-wave field from transversal motion}

Consider, first of all, the Klein--Gordon equation with a plane-wave field
(\ref{pw.22}). We shall be interested in such solutions of this equation that
are eigenfunctions of the operator $\hat{\lambda}=\left(  n\hat{p}\right)  $,
\begin{equation}
\hat{\lambda}\varphi_{\lambda}\left(  x\right)  =\lambda\varphi_{\lambda
}\left(  x\right)  \,. \label{pw.45}%
\end{equation}
However, as was already mentioned, we do not demand that the wave functions
$\varphi_{\lambda}\left(  x\right)  $ be eigenfunctions of the operator
$\mathbf{\hat{p}}_{\perp}.$ The general solution of equation (\ref{pw.45}) is%
\begin{equation}
\varphi_{\lambda}\left(  x\right)  =\exp\left(  -i\lambda x^{0}+i\frac{\lambda
}{2}\xi\right)  \Phi_{\lambda}\left(  \xi,\mathbf{r}_{\bot}\right)
\,,\;\mathbf{r}_{\bot}=\left(  x^{1},x^{2},0\right)  \,, \label{pw.46}%
\end{equation}
where $\Phi_{\lambda}\left(  \xi,\mathbf{r}_{\bot}\right)  $ is an arbitrary
function of the indicated arguments. This function has to obey the equation%
\begin{equation}
i\frac{\partial\Phi_{\lambda}}{\partial\xi}=\hat{H}\Phi_{\lambda
}\,,\;\hat{H}=\frac{1}{2\lambda}\left(  \mathbf{\hat{P}}_{\bot}^{2}%
+m^{2}\right)  \,,\;\mathbf{\hat{P}}_{\bot}^{2}=\mathbf{\hat{p}}_{\perp}%
^{2}-e\mathbf{A}\left(  \xi\right)  \,. \label{pw.47}%
\end{equation}
Equation (\ref{pw.47}) is a nonstationary two-dimensional Schr\"{o}dinger
equation with respect to the ``time'' $\xi.$

We can see that there exists a transformation that eliminates the plane-wave
field $\mathbf{A}\left(  \xi\right)  $ from equation (\ref{pw.47}) and reduces
the latter to a free two-dimensional Schr\"{o}dinger equation. The
transformation consists of a change of variables $\mathbf{r}_{\bot}%
\rightarrow\mathbf{x}_{\bot}$,\
\begin{equation}
\mathbf{r}_{\bot}=\mathbf{x}_{\bot}-\frac{e}{\lambda}\int\mathbf{A}\left(
\xi\right)  d\xi\,,\label{pw.48}%
\end{equation}
and a function replacement, $\Phi_{\lambda}\left(  \xi,\mathbf{r}_{\bot
}\right)  \rightarrow\Psi_{\lambda}\left(  \xi,\mathbf{r}_{\bot}\right)  ,$%
\begin{align}
&  \Phi_{\lambda}\left(  \xi,\mathbf{r}_{\bot}\right)  =\exp\left[
-i\alpha\left(  \xi\right)  \right]  \Psi_{\lambda}\left(  \xi,\mathbf{r}%
_{\bot}\mathbf{+}\frac{e}{\lambda}\int\mathbf{A}\left(  \xi\right)
d\xi\right)  \,,\nonumber\\
&  \alpha\left(  \xi\right)  =\frac{1}{2\lambda}\int\left[  e^{2}%
\mathbf{A}^{2}\left(  \xi\right)  +m^{2}\right]  d\xi\,.\label{pw.49a}%
\end{align}
It is a simple task to verify that the function $\Psi_{\lambda}\left(
\xi,\mathbf{r}_{\bot}\right)  $ is a solution of a free two-dimensional
Schr\"{o}dinger equation of the form%
\begin{equation}
i\frac{\partial\Psi_{\lambda}\left(  \xi,\mathbf{r}_{\bot}\right)  }%
{\partial\xi}=\hat{H}_{0}\Psi_{\lambda}\left(  \xi,\mathbf{r}_{\bot}\right)
\,,\;\hat{H}_{0}=\frac{\mathbf{\hat{p}}_{\perp}^{2}}{2\lambda}\,.\label{pw.50}%
\end{equation}

Finally, we have a set of solutions to the Klein--Gordon equation with a
plane-wave field in the form%
\begin{align}
&  \varphi_{\lambda}\left(  x\right)  =\exp\left(  -i\Gamma\right)
\Psi_{\lambda}\left(  \xi,\mathbf{r}_{\bot}\mathbf{+}\frac{e}{\lambda}%
\int\mathbf{A}\left(  \xi\right)  d\xi\right)  \,,\nonumber\\
&  \Gamma=\lambda x^{0}-\frac{\lambda}{2}\xi+\frac{1}{2\lambda}\int\left[
e^{2}\mathbf{A}^{2}\left(  \xi\right)  +m^{2}\right]  d\xi\,,\label{pw.51}%
\end{align}
where the function $\Psi_{\lambda}\left(  \xi,\mathbf{r}_{\bot}\right)  $ is a
solution of the free two-dimensional Schr\"{o}dinger equation (\ref{pw.50})
that does not contain the plane-wave field.

It is interesting to note that the change of variables (\ref{pw.48}) also
eliminates the plane-wave field from the classical equations of motion.
Indeed, this follows from the Lorentz equations%
\begin{equation}
m\mathbf{\ddot{r}}_{\bot}=-e\mathbf{A}^{\prime}\left(  \xi\right)  \dot{\xi
}\,. \label{pw.52}%
\end{equation}
In terms of the variables $\mathbf{x}_{\bot}$, related to $\mathbf{r}_{\bot}$
by (\ref{pw.48}), we have the equations of free motion $m\mathbf{\ddot{x}%
}_{\bot}=0.$

Similarly, we find that there exist solutions of the Dirac equation with a
plane-wave field in the form%
\begin{equation}
\psi_{\lambda}\left(  x\right)  =\exp\left(  -i\Gamma\right)  \hat{R}%
\Psi_{\lambda}\left(  \xi,\mathbf{r_{\bot}+}\frac{e}{\lambda}\int
\mathbf{A}\left(  \xi\right)  d\xi\right)  \vartheta\,,\label{pw.53}%
\end{equation}
where $\vartheta$ is an arbitrary two-component spinor, while the operator
$\hat{R}$ is
\begin{equation}
\hat{R}=\left(
\begin{array}
[c]{c}%
m+\lambda+\sigma_{3}\mathbf{\sigma\hat{P}}_{\bot}\\
\left(  m-\lambda\right)  \sigma_{3}+\mathbf{\sigma\hat{P}}_{\bot}%
\end{array}
\right)  ,\label{pw.54}%
\end{equation}
and the function $\Psi_{\lambda}\left(  \xi,\mathbf{r}_{\bot}\right)  $ is a
solution of the free two-dimensional Schr\"{o}dinger equation (\ref{pw.50}).

\subsection{Examples of non-Volkov solutions}

We now present some specific examples of non-Volkov solutions, considering
different solutions of Eq. (\ref{pw.50}).

\subsubsection{First example}

Let us consider Eq. (\ref{pw.50}). We now introduce the dimensionless
variables $r_{s}$, $s=1,2\,,$ and $\tau$ that are related to the variables
$x^{s}$ and $\xi$ as follows:%
\begin{equation}
\eta_{s}=2\lambda x^{s}+2e\int A^{s}\left(  \xi\right)  d\xi\,,\;\tau
=2\lambda\xi\,,\;A^{1}\left(  \xi\right)  =A_{x}\left(  \xi\right)
,\;A^{2}\left(  \xi\right)  =A_{y}\left(  \xi\right)  \,.\label{20}%
\end{equation}
In terms of the new variables, we have%
\begin{align}
&  \hat{K}\Psi\left(  \eta_{1},\eta_{2},\tau\right)  =0\,,\;\hat
{K}=i\frac{\partial}{\partial\tau}+\frac{\partial^{2}}{\partial\eta_{1}^{2}%
}+\frac{\partial^{2}}{\partial\eta_{2}^{2}}\,,\nonumber\\
&  \Psi_{\lambda}\left(  \xi,\mathbf{r}_{\bot}\right)  =\Psi\left(  \eta
_{1},\eta_{2},\tau\right)  \,.\label{21}%
\end{align}
We introduce the creation and annihilation operators as%
\begin{align}
&  a_{s}=\frac{1}{\sqrt{2}}\left(  \eta_{s}+\frac{\partial}{\partial\eta_{s}%
}\right)  \,,\;a_{s}^{+}=\frac{1}{\sqrt{2}}\left(  \eta_{s}-\frac{\partial
}{\partial\eta_{s}}\right)  \,,\label{22}\\
&  \left[  a_{s},a_{s^{\prime}}\right]  =\left[  a_{s}^{+},a_{s^{\prime}}%
^{+}\right]  =0\,,\;\left[  a_{s},a_{s^{\prime}}^{+}\right]  =\delta
_{ss^{\prime}}\,,\;s=1,2\,.\label{23}%
\end{align}
In terms of these operators,%
\begin{equation}
\hat{K}=i\frac{\partial}{\partial\tau}-\sum_{s=1,2}H_{s}\,,\;2H_{s}=a_{s}%
a_{s}^{+}+a_{s}^{+}a_{s}-a_{s}^{2}-a_{s}^{+2}\,.\label{24}%
\end{equation}

Let us construct such integrals of motion $A_{s}$ for equation (\ref{21}) that
are linear in the creation and annihilation operators of the same kind (the
same $s$),%
\begin{align}
&  A_{s}=f_{s}\left(  \tau\right)  a_{s}+g_{s}\left(  \tau\right)  a_{s}%
^{+}\,,\label{25}\\
&  \left[  A_{s},\hat{K}\right]  =0\,,\;s=1,2\,. \label{26}%
\end{align}
It follows from (\ref{26}) that the functions $f_{s}$ and $g_{s}$ obey the
equations
\begin{equation}
i\dot{f}_{s}+f_{s}+g_{s}=0\,,\;i\dot{g}_{s}-f_{s}-g_{s}=0\,. \label{27}%
\end{equation}
The general solution of the set (\ref{27}) has the form%
\begin{equation}
f_{s}\left(  \tau\right)  =c_{1}^{\left(  s\right)  }+i\left(  c_{1}^{\left(
s\right)  }+c_{2}^{\left(  s\right)  }\right)  \tau\,,\;g_{s}\left(
\tau\right)  =c_{2}^{\left(  s\right)  }-i\left(  c_{1}^{\left(  s\right)
}+c_{2}^{\left(  s\right)  }\right)  \tau\,,\;s=1,2\,, \label{28}%
\end{equation}
where $c_{1}^{\left(  s\right)  }$ and $c_{2}^{\left(  s\right)  }$ are
arbitrary complex numbers. One can easily check (taking (\ref{23}) into
account) that the above-introduced integrals of motion obey the following
commutation relations:%
\begin{align}
&  \left[  A_{s},A_{s^{\prime}}\right]  =\left[  A_{s}^{+},A_{s^{\prime}}%
^{+}\right]  =0\,,\;\left[  A_{s},A_{s^{\prime}}^{+}\right]  =\Delta_{s}%
\delta_{ss^{\prime}}\,,\nonumber\\
&  \Delta_{s}=\left|  f_{s}\right|  ^{2}-\left|  g_{s}\right|  ^{2}=\left|
c_{1}^{\left(  s\right)  }\right|  ^{2}-\left|  c_{2}^{\left(  s\right)
}\right|  ^{2}\,,\;s=1,2\,. \label{29}%
\end{align}

Since $A_{s}$ are integrals of motion, we can look for such solutions of
equation (\ref{21}) that are eigenvectors of $A_{s},$%
\begin{equation}
A_{s}\Psi_{z}\left(  \eta_{1},\eta_{2},\tau\right)  =z_{s}\Psi_{z}\left(
\eta_{1},\eta_{2},\tau\right)  \,,\;s=1,2\,.\label{30}%
\end{equation}
Here, $z_{s}$ are arbitrary complex numbers. Such solutions can be chosen as%
\begin{equation}
\Psi_{z_{1}z_{2}}\left(  \eta_{1},\eta_{2},\tau\right)  =\Psi_{z_{1}}\left(
\eta_{1},\tau\right)  \Psi_{z_{2}}\left(  \eta_{2},\tau\right)  \,,\label{31}%
\end{equation}
where the function $\Psi_{z_{s}}\left(  \eta_{s},\tau\right)  $ obeys similar
(for each $s$) equations:%
\begin{align}
&  \left(  i\frac{\partial}{\partial\tau}-H_{s}\right)  \Psi_{z_{s}}\left(
\eta_{s},\tau\right)  =0\,,\label{32a}\\
&  A_{s}\Psi_{z_{s}}\left(  \eta_{s},\tau\right)  =z_{s}\Psi_{z_{s}}\left(
\eta_{s},\tau\right)  \,,\;s=1,2\,.\label{32}%
\end{align}
In what follows, we analyze these equations for a fixed $s$ ($s$ is then
omitted in all the quantities).

We first consider the case $\Delta=0.$ In this case, the operator $A$ can be
chosen, without loss of generality, as self-adjoint, $A^{+}=A$. Then
$z=z^{\ast},\;g=f^{\ast},\;f\left(  \tau\right)  =c+i\left(  c+c^{\ast
}\right)  \tau\,,$ and solutions of the set (\ref{32a}), (\ref{32}) have the
form%
\begin{equation}
\Psi_{z}\left(  \eta,\tau\right)  =\left[  \sqrt{2}\pi\left(  f-f^{\ast
}\right)  \right]  ^{-1/2}\exp\left[  \frac{f+f^{\ast}}{2\left(  f^{\ast
}-f\right)  }\left(  \eta-\frac{\sqrt{2}z}{f+f^{\ast}}\right)  ^{2}\right]
\,.\label{33}%
\end{equation}
The functions (\ref{33}) obey the following relations of orthogonality and
completeness:
\begin{equation}
\int_{-\infty}^{\infty}\Psi_{z}^{\ast}\left(  \eta,\tau\right)  \Psi
_{z^{\prime}}\left(  \eta,\tau\right)  d\eta=\delta\left(  z-z^{\prime
}\right)  \,,\;\int_{-\infty}^{\infty}\Psi_{z}^{\ast}\left(  \eta^{\prime
},\tau\right)  \Psi_{z}\left(  \eta,\tau\right)  dz=\delta\left(  \eta
-\eta^{\prime}\right)  \,.\label{34}%
\end{equation}

If $\Delta>0,$ then, without loss of generality, we can choose $\Delta
=1$,\thinspace multiplying $A$ by a complex number. In this case, $A$ and
$A^{+}$ are annihilation and creation operators, and solutions $\Psi_{\lambda
z_{s}}$ of equation (\ref{32}) are coherent states. In order to obey equation
(\ref{32a}), these states must take the form%
\begin{align}
&  \Psi_{z}\left(  \eta,\tau\right)  =\left(  f-g\right)  ^{-1/2}U_{0}\left(
q\right)  \exp\Theta\,,\nonumber\\
&  q=\left[  2\eta-\sqrt{2}z\left(  f^{\ast}-g^{\ast}\right)  -\sqrt{2}%
z^{\ast}\left(  f-g\right)  \right]  \left[  2\left|  f-g\right|  \right]
^{-1}\,,\nonumber\\
&  \Theta=\left\{  2\eta^{2}\left(  fg^{\ast}-gf^{\ast}\right)  +2\sqrt{2}%
\eta\left[  z\left(  f^{\ast}-g^{\ast}\right)  -z^{\ast}\left(  f-g\right)
\right]  \right.  \nonumber\\
&  \left.  +z^{\ast2}\left(  f-g\right)  ^{2}-z^{2}\left(  f^{\ast}-g^{\ast
}\right)  ^{2}\right\}  \left[  4\left|  f-g\right|  \right]  ^{-1}%
\,.\label{35}%
\end{align}
Here, $U_{0}\left(  x\right)  $ is the zero function from the set of Hermite
functions $U_{n}\left(  x\right)  =\left(  2^{n}n!\sqrt{\pi}\right)
^{-1/2}\exp\left(  -x^{2}/2\right)  H_{n}\left(  x\right)  ,$ where
$H_{n}\left(  x\right)  $ are Hermite polynomials \cite{14}.

Finally, if $\Delta<0,$ then the operator $A$ has no eigenfunctions that can
be normalized, even as distributions. However, in this case the operator
$B=A^{+}$ is indeed an annihilation operator, and the above consideration is
applicable here.

Having the expressions for the functions $\Psi_{z}\left(  \eta_{1},\eta
_{2},\tau\right)  ,$ we can construct the corresponding solutions of the
Klein--Gordon equation (\ref{pw.22}) and of the Dirac equation (\ref{pw.26}),
with the help of formulae (\ref{pw.51}) and (\ref{pw.53}). Such solutions are
eigenfunctions of the integrals of motion $\mathcal{A}_{s}$ $,$ $s=1,2,$ which
can be constructed from the operators $A_{s}$, with allowance for the
transformation (\ref{20}),%
\begin{align}
&  \mathcal{A}_{s}=\frac{f_{s}\left(  2\hat{\lambda}\xi\right)  }{\sqrt{2}%
}\left[  2\hat{\lambda}x^{s}+2e\int A^{s}\left(  \xi\right)  d\xi
\,+\frac{1}{2\hat{\lambda}}\frac{\partial}{\partial x^{s}}\right]  \nonumber\\
&  +\frac{g_{s}\left(  2\hat{\lambda}\xi\right)  }{\sqrt{2}}\left[
2\hat{\lambda}x^{s}+2e\int A^{s}\left(  \xi\right)  d\xi\,-\frac{1}%
{2\hat{\lambda}}\frac{\partial}{\partial x^{s}}\right]  \,,\nonumber\\
&  \left[  \mathcal{A}_{s},\mathcal{A}_{s^{\prime}}\right]  =\left[
\mathcal{A}_{s}^{+},\mathcal{A}_{s^{\prime}}^{+}\right]  =0\,,\;\left[
\mathcal{A}_{s},\mathcal{A}_{s^{\prime}}^{+}\right]  =\Delta_{s}%
\delta_{ss^{\prime}}\,,\nonumber\\
&  \left[  \mathcal{A}_{s},\mathcal{\hat{K}}\right]  =\left[  \mathcal{A}%
_{s}^{+},\mathcal{\hat{K}}\right]  =0\,,\;\left[  \mathcal{A}_{s}%
,\mathcal{\hat{D}}\right]  =\left[  \mathcal{A}_{s}^{+},\mathcal{\hat{D}%
}\right]  =0\,.\label{39}%
\end{align}
For example, in the case $\Delta_{s}=1,$ $s=1,2,$ we obtain squeezed coherent
states describing the transversal motion of a charge in a plane-wave field
(transversal squeezed coherent states). The squeezing of the states is
determined by the variation of the constants $c$ in formulae (\ref{28}). One
can easily verify that in such a case the mean values of the transversal
coordinates obey the classical equations of motion (see the following example).

\subsubsection{Second example}

Consider the case $\Delta=1.$ Since the operator $A^{+}$ is an integral of
motion, we can construct solutions of equation (\ref{32a}) (which no longer
obey Eq. (\ref{33})) as follows:%
\begin{equation}
\Psi_{z,n}\left(  \eta,\tau\right)  =\left(  A^{+}-z^{\ast}\right)  ^{n}%
\Psi_{z}\left(  \eta,\tau\right)  \,,\;n=0,1,2,...\;.\label{36}%
\end{equation}
For $n=0$, they coincide with the solutions considered in the first example.
The functions (\ref{36}) may be called generalized squeezed coherent states.
They have the explicit form%
\begin{equation}
\Psi_{z,n}\left(  \eta,\tau\right)  =\left(  f-g\right)  ^{-1/2}\left(
\frac{f^{\ast}-g^{\ast}}{f-g}\right)  ^{\frac{n}{2}}U_{n}\left(  q\right)
\exp\Theta\,,\label{37}%
\end{equation}
and the following properties:%
\begin{align}
&  \left(  A-z\right)  \Psi_{z,n}=\sqrt{n}\Psi_{z,n-1}\,,\nonumber\\
&  \left(  A^{+}-z^{\ast}\right)  \Psi_{z,n}=\sqrt{n+1}\Psi_{z,n+1}%
\,,\nonumber\\
&  \int_{-\infty}^{\infty}\Psi_{z,n}^{\ast}\left(  \eta,\tau\right)
\Psi_{z,n^{\prime}}\left(  \eta,\tau\right)  d\eta=\delta_{nn^{\prime}%
}\,,\nonumber\\
&  \sum_{n=0}^{\infty}\Psi_{z,n}^{\ast}\left(  \eta^{\prime},\tau\right)
\Psi_{z,n}\left(  \eta,\tau\right)  =\delta\left(  \eta-\eta^{\prime}\right)
\nonumber\\
&  \int_{-\infty}^{\infty}\Psi_{z,n}^{\ast}\left(  \eta^{\prime},\tau\right)
\Psi_{z,n^{\prime}}\left(  \eta,\tau\right)  d^{2}z=\pi\delta_{nn^{\prime}%
}\delta\left(  \eta-\eta^{\prime}\right)  \,.\label{38}%
\end{align}

Using (\ref{37}), we can now construct solutions of the Klein--Gordon and
Dirac equations by means of formulae (\ref{pw.51}) and (\ref{pw.53})
(transversal generalized squeezed coherent states). Calculating the mean
values of the operators $\mathbf{\hat{r}}_{\perp}$and\ $\mathbf{\hat{p}%
}_{\perp}$ on such states, we obtain (the results are the same for both the
Klein--Gordon and the Dirac solutions, and do not depend on the quantum number
$n$)%
\begin{equation}
\left\langle \mathbf{\hat{r}}_{\perp}\right\rangle =\mathbf{r}_{\perp}\left(
0\right)  +\lambda^{-1}\int_{0}^{\xi}\mathbf{P}_{\perp}\left(  \xi\right)
d\xi\,,\;\left\langle \mathbf{\hat{p}}_{\perp}\right\rangle =\mathbf{p}%
_{\perp}\,,\label{41}%
\end{equation}
where $\mathbf{r}_{\perp}\left(  0\right)  =\left(  x_{0}^{1},x_{0}%
^{2},0\right)  \,,\;\mathbf{p}_{\perp}=\left(  p^{1},p^{2},0\right)  ,$ and%
\begin{align}
x_{0}^{s} &  =\left(  2\sqrt{2}\lambda\right)  ^{-1}\left[  \left(
c_{1}^{s\ast}-c_{2}^{s\ast}\right)  z_{s}+\left(  c_{1}^{s}-c_{2}^{s}\right)
z_{s}^{\ast}\right]  \,,\nonumber\\
p^{s} &  =i\sqrt{2}\lambda\left[  \left(  c_{1}^{s}+c_{2}^{s}\right)
z_{s}^{\ast}-\left(  c_{1}^{s\ast}+c_{2}^{s\ast}\right)  z_{s}\right]
\,,\;s=1,2\,.\label{43}%
\end{align}
It is easy to see from (\ref{pw.21}) and (\ref{pw.52}) that the mean values
(\ref{41}) follow the classical trajectory of transversal motion with the
initial data (at $\xi=0$) given by $\mathbf{r}_{\perp}\left(  0\right)  .$ To
provide such initial data, one has to select states with%
\[
z_{s}=\sqrt{2}\lambda x_{0}^{s}\left(  c_{1}^{s}+c_{2}^{s}\right)
+\frac{ip^{s}}{2\sqrt{2}\lambda}\left(  c_{1}^{s}-c_{2}^{s}\right)  \,.
\]

\subsubsection{Third example}

Consider the operator%
\begin{align}
&  \hat{L}_{z}=i\left(  \eta_{2}\frac{\partial}{\partial\eta_{1}}-\eta
_{1}\frac{\partial}{\partial\eta_{2}}\right)  =-i\frac{\partial}%
{\partial\varphi}\,,\nonumber\\
&  \eta_{1}=\rho\cos\varphi\,,\;\eta_{2}=\rho\sin\varphi\,,\;\rho=\sqrt
{\eta_{1}^{2}+\eta_{2}^{2}}\,, \label{45}%
\end{align}
which is an integral of motion for equation (\ref{21})%
\[
\left[  \hat{L}_{z},\hat{K}\right]  =0\,.
\]
It can be interpreted as the $z$-projection of the angular momentum operator.
The corresponding integral of motion for the Klein--Gordon equation can be
easily recovered. It has the form%
\begin{equation}
\mathcal{\hat{L}}_{z}=\left(  \left[  \mathbf{r}_{\bot}+\frac{e}{\hat{\lambda
}}\int\mathbf{A}\left(  \xi\right)  d\xi\,,\mathbf{\hat{p}}_{\bot}\right]
\mathbf{k}\right)  \,,\;\left[  \mathcal{\hat{L}}_{z},\mathcal{\hat{K}%
}\right]  =0\,, \label{46}%
\end{equation}
where $\mathbf{k}$ is a unit vector in the $z$-direction.

Let us seek for such solutions of equation (\ref{21}) that are eigenvectors of
$\hat{L}_{z},$%

\begin{equation}
\hat{L}_{z}\Psi_{l}\left(  \eta_{1},\eta_{2},\tau\right)  =l\Psi_{l}\left(
\eta_{1},\eta_{2},\tau\right)  \,,\;l=0,\pm1,\pm2,...\;. \label{47}%
\end{equation}
The functions $\Psi_{l}\left(  \eta_{1},\eta_{2},\tau\right)  $ have the form%
\begin{equation}
\Psi_{l}\left(  \eta_{1},\eta_{2},\tau\right)  =\chi_{l}\left(  \rho
,\tau\right)  \exp il\varphi\;,\;\left(  \exp i\varphi=\left(  \eta_{1}%
+i\eta_{2}\right)  /\rho\right)  \,, \label{48}%
\end{equation}
where the functions $\chi_{l}\left(  \rho,\xi\right)  $ obey the equation%
\begin{equation}
\left(  i\frac{\partial}{\partial\tau}+\frac{\partial^{2}}{\partial\rho^{2}%
}+\frac{1}{\rho}\frac{\partial}{\partial\rho}-\frac{l^{2}}{\rho^{2}}\right)
\chi_{l}\left(  \rho,\tau\right)  =0 \label{49}%
\end{equation}
A general solution of the latter equation, with the initial condition
$\chi_{l}\left(  \rho,0\right)  =\chi_{l}^{\left(  0\right)  }\left(
\rho\right)  $, has the form%
\begin{align}
&  \chi_{l}\left(  \rho,\tau\right)  =\int_{0}^{\infty}G_{l}\left(  \rho
,\rho^{\prime};\tau\right)  \chi_{l}^{\left(  0\right)  }\left(  \rho^{\prime
}\right)  d\rho^{\prime}\,,\nonumber\\
&  G_{l}\left(  \rho,\rho^{\prime};\tau\right)  =\frac{\left(  -i\right)
^{l+1}\rho^{\prime}}{2\tau}J_{l}\left(  \frac{\rho\rho^{\prime}}{2\tau
}\right)  \exp\left(  i\frac{\rho^{2}+\rho^{\prime2}}{4\tau}\right)  \,,
\label{50}%
\end{align}
where $J_{l}\left(  x\right)  $ are Bessel functions \cite{14}. One can
consider stationary states, choosing $\chi_{l}\left(  \rho,\tau\right)
=f_{lp}\left(  \rho\right)  \exp\left(  -ip\tau\right)  .$ Then$\;f_{lp}%
\left(  \rho\right)  =J_{l}\left(  \sqrt{p}\rho\right)  .$

The corresponding solutions of the Klein--Gordon equations can be easily found
with the help of formulae (\ref{pw.51}). Switching on the plane-wave field in
these solutions, we obtain the states of a free relativistic particle with a
definite $z$-projection of the angular momentum.

\section{Combined field}

Consider now a combination of a plane-wave field and colinear constant
electric and magnetic fields. The covariant components of the corresponding
electromagnetic potentials are chosen as%
\begin{equation}
A_{0}=A_{3}=A_{0}\left(  \xi\right)  \,,\;A_{1}=\frac{H}{2}x^{2}+A_{1}\left(
\xi\right)  \,,\;A_{2}=-\frac{H}{2}x^{1}+A_{2}\left(  \xi\right)  \,,
\label{cf.1}%
\end{equation}
where $A_{k}\left(  \xi\right)  ,\;k=1,2,$ are arbitrary functions of $\xi,$
and $H=\mathrm{const}.$ The corresponding components of the electric and
magnetic fields are%
\begin{equation}
E_{x}=H_{y}=A_{1}^{\prime}\left(  \xi\right)  ,\;E_{y}=-H_{x}=A_{2}^{\prime
}\left(  \xi\right)  \,,\;E_{z}=2A_{0}^{\prime}\left(  \xi\right)
\,,\;H_{z}=H. \label{cf.2}%
\end{equation}
In what follows, we call the electromagnetic field (\ref{cf.2}) the combined
electromagnetic field.

The Lorentz equations with the combined electromagnetic field can be written
as%
\begin{align}
m\ddot{x}^{0} &  =e\left(  A_{1}^{\prime}\dot{x}^{1}+A_{2}^{\prime}\dot{x}%
^{2}+A_{0}^{\prime}\dot{x}^{3}\right)  ,\;m\ddot{x}^{1}=e\left(  A_{1}%
^{\prime}\dot{\xi}+H\dot{x}^{2}\right)  \,,\nonumber\\
m\ddot{x}^{2} &  =e\left(  A_{2}^{\prime}\dot{\xi}-H\dot{x}^{1}\right)
\,,\;m\ddot{x}^{3}=e\left(  A_{1}^{\prime}\dot{x}^{1}+A_{2}^{\prime}\dot
{x}^{2}+2A_{0}^{\prime}\dot{x}^{0}\right)  \,.\label{cf.3}%
\end{align}
Dots stand for derivatives with respect to the proper time $\tau$. Solutions
of equations (\ref{cf.3}) are known, see \cite{13}. We only remark here that
these equations imply the conservation of the quantity $\lambda,$%
\begin{align}
\lambda &  =m\left(  \dot{x}^{0}-\dot{x}^{3}\right)  +2eA_{0}\left(
\xi\right)  =m\dot{\xi}+2eA_{0}\left(  \xi\right)  \nonumber\\
&  =P_{0}+P_{3}+2eA_{0}\left(  \xi\right)  =p_{0}+p_{3}\,.\label{cf.4}%
\end{align}
Here, $p_{\mu}$ are components of the generalized momentum, and $P_{\mu}%
=m\dot{x}_{\mu}=p_{\mu}-eA_{\mu}$ are components of the kinetic momentum. We
introduce the notation%
\begin{align}
&  2eA_{0}\left(  \xi\right)  =-a_{0}\left(  \xi\right)  \,,\;eA_{k}\left(
\xi\right)  =a_{k}\left(  \xi\right)  \,,\nonumber\\
&  eH=\gamma\,,\;\mathcal{P}=\mathcal{P}\left(  \xi\right)  =\lambda
+a_{0}\left(  \xi\right)  \,.\label{cf.6}%
\end{align}
Then (\ref{cf.4}) takes the form%
\begin{equation}
m\dot{\xi}=\mathcal{P}\left(  \xi\right)  =\lambda+a_{0}\left(  \xi\right)
\,.\label{cf.7}%
\end{equation}
Integrating (\ref{cf.7}), we obtain a relation between the proper time $\tau$
and the variable $\xi,$%
\begin{equation}
\tau\left(  \xi\right)  =m\int\frac{d\xi}{\mathcal{P}\left(  \xi\right)
}\,.\label{cf.8}%
\end{equation}

The Lorentz equations for $x^{k},\;k=1,2,$ take the form%
\begin{equation}
m\ddot{x}^{1}=a_{1}^{\prime}\left(  \xi\right)  \dot{\xi}+\gamma\dot{x}%
^{2}\,,\;m\ddot{x}^{2}=a_{2}^{\prime}\left(  \xi\right)  \dot{\xi}-\gamma
\dot{x}^{1}\,.\label{cf.9}%
\end{equation}
in terms of notation (\ref{cf.6}). We make the change of variables%
\begin{equation}
x^{1}=X-q_{1}\left(  \xi\right)  \,,\;x^{2}=Y-q_{2}\left(  \xi\right)
\,,\label{cf.10}%
\end{equation}
where the real functions $q_{k}\left(  \xi\right)  ,\;k=1,2$, are defined as%
\begin{equation}
q\left(  \xi\right)  =q_{1}\left(  \xi\right)  +iq_{2}\left(  \xi\right)
=-\exp\left[  -\frac{i\gamma\tau\left(  \xi\right)  }{m}\right]  \int
\exp\left[  \frac{i\gamma\tau\left(  \xi\right)  }{m}\right]  \frac{a\left(
\xi\right)  }{\mathcal{P}\left(  \xi\right)  }d\xi\,,\label{cf.11}%
\end{equation}
with $a\left(  \xi\right)  =a_{1}\left(  \xi\right)  +ia_{2}\left(
\xi\right)  .$ The complex function $q\left(  \xi\right)  $ obeys the equation%
\begin{equation}
\mathcal{P}\left(  \xi\right)  q^{\prime}\left(  \xi\right)  +i\gamma q\left(
\xi\right)  +a\left(  \xi\right)  =0\,.\label{cf.12}%
\end{equation}
In terms of the new variables $u$ and $v$, equations (\ref{cf.9}) take the
form%
\begin{equation}
m\ddot{X}-\gamma\dot{Y}=0\,,\;m\ddot{Y}+\gamma\dot{X}=0\,.\label{cf.13}%
\end{equation}
The set (\ref{cf.13}) contains neither the plane-wave field nor the colinear
electric field. It describes the motion of a charged particle in a constant
magnetic field in the plane $x^{3}=\mathrm{const.}$ The above consideration
indicates that the plane-wave field and the colinear electric field can be
eliminated from the quantum equations of motion as well.

We now consider the Klein--Gordon and Dirac equations with the combined
electromagnetic field. Exact solutions of these equations were first found in
\cite{12,13}. As in the case of a plane-wave, we present below new classes of
solutions of these equations. To this end, we are going to represent a
transformation that eliminates the plane-wave field and the colinear electric
field from the transversal motion. Thus, the transversal motion in the
combined electromagnetic field is mapped to the nonrelativistic transversal
motion in the constant uniform magnetic field.

In quantum theory, the operator $\widehat{\lambda}$ corresponding to the
classical quantity (\ref{cf.4}) is an integral of motion. We seek solutions of
the Klein--Gordon equation as eigenfunctions of the operator $\widehat
{\lambda}$. Such solutions have the form%
\[
\varphi_{\lambda}\left(  x\right)  =\frac{1}{\sqrt{\mathcal{P}\left(
\xi\right)  }}\exp\left(  -i\lambda x^{0}+i\frac{\lambda}{2}\xi\right)
\Phi_{\lambda}\left(  \xi,x^{1},x^{2}\right)  \,,
\]
where the function $\Phi_{\lambda}\left(  \xi,x^{1},x^{2}\right)  $ has to
obey the equation%
\begin{align}
&  2i\mathcal{P}\left(  \xi\right)  \frac{\partial\Phi_{\lambda}}{\partial\xi
}=\left(  \hat{P}_{1}^{2}+\hat{P}_{2}^{2}+m^{2}\right)  \Phi_{\lambda
}\,,\nonumber\\
&  \hat{P}_{k}=i\partial_{k}-eA_{k}\left(  \xi\right)
,\;k=1,2\,.\label{cf.20}%
\end{align}
We now use the variables $X,Y$ and $\tau\left(  \xi\right)  $, see
(\ref{cf.8}) and (\ref{cf.10}), and a function replacement, $\Phi_{\lambda
}\rightarrow\Psi_{\lambda}$, for eliminating the plane-wave field and the
colinear electric field from the equations,%
\begin{align}
&  \Phi_{\lambda}\left(  \xi,x^{1},x^{2}\right)  =\exp\left(  -i\Gamma\right)
\Psi_{\lambda}\left(  \tau,X,Y\right)  \,,\nonumber\\
&  \Gamma=\frac{i\gamma}{4}\left[  q^{\ast}\left(  \xi\right)  \left(
x^{1}+ix^{2}\right)  -q\left(  \xi\right)  \left(  x^{1}-ix^{2}\right)
\right]  \nonumber\\
&  +\int\frac{d\xi}{2\mathcal{P}\left(  \xi\right)  }\left\{  \left|  a\left(
\xi\right)  \right|  ^{2}+m^{2}+\frac{i\gamma}{2}\left[  q\left(  \xi\right)
a^{\ast}\left(  \xi\right)  -q^{\ast}\left(  \xi\right)  a\left(  \xi\right)
\right]  \right\}  \,.\label{cf.21}%
\end{align}
The function $\Psi_{\lambda}\left(  \tau,X,Y\right)  $ obeys the equation%
\begin{align}
&  i\frac{\partial\Psi_{\lambda}\left(  \tau,X,Y\right)  }{\partial\tau
}=\frac{\left(  \hat{\pi}_{1}^{2}+\hat{\pi}_{2}^{2}\right)  }{2m}\Psi
_{\lambda}\left(  \tau,X,Y\right)  \,,\nonumber\\
&  \hat{\pi}_{1}=i\partial_{X}-\frac{\gamma}{2}Y\,,\;\hat{\pi}_{2}%
=i\partial_{Y}+\frac{\gamma}{2}X\,.\label{cf.14}%
\end{align}
This is the two-dimensional Schr\"{o}dinger equation for a charged particle in
a constant uniform magnetic field. In this equation, the plane-wave field and
the colinear electric field are already eliminated.

We now pass to the Dirac equation with the combined electromagnetic field. We
shall seek solutions of the Dirac equation as eigenfunctions of the operator
$\widehat{\lambda}.$ Such solutions have the form%
\begin{equation}
\psi_{\lambda}\left(  x\right)  =\exp\left(  -i\lambda x^{0}+i\frac{\lambda
}{2}\xi\right)  \left(
\begin{array}
[c]{c}%
V+U\\
\sigma_{3}\left(  V-U\right)
\end{array}
\right)  \,.\label{cf.15}%
\end{equation}
Here, $V=V\left(  \xi,x^{1},x^{2}\right)  $ and $U=U\left(  \xi,x^{1}%
,x^{2}\right)  $ are two-component spinors that have to obey the equations%
\begin{align}
&  2i\frac{\partial U}{\partial\xi}=\left[  m+\left(  \mathbf{\sigma
}\,\mathbf{\hat{P}}_{\bot}\right)  \sigma_{3}\right]  V\,,\nonumber\\
&  V=\mathcal{P}^{-1}\left(  \xi\right)  \left[  m-\left(  \mathbf{\sigma
}\,\mathbf{\hat{P}}_{\bot}\right)  \sigma_{3}\right]  U\,,\label{cf.16}%
\end{align}
where $\mathbf{\hat{P}}_{\bot}=-\left(  \hat{P}_{1},\hat{P}_{2},0\right)  .$
Obviously, it is sufficient to know the spinor $U.$ The latter spinor obeys
the equation%
\begin{equation}
2i\mathcal{P}\left(  \xi\right)  \frac{\partial U}{\partial\xi}=\left(
\mathbf{\hat{P}}_{\bot}^{2}+m^{2}-\gamma\sigma_{3}\right)  U\,.\label{cf.17}%
\end{equation}
We transform this equation by the spinor replacement $U\rightarrow\Theta,$%
\[
U\left(  \xi,x^{1},x^{2}\right)  =\exp\left[  i\frac{\gamma\tau\left(
\xi\right)  \sigma_{3}}{2m}\right]  \Theta\left(  \xi,x^{1},x^{2}\right)  \,.
\]
The new spinor $\Theta\left(  \xi,x^{1},x^{2}\right)  $ has to obey the
following equation:%
\begin{equation}
2i\mathcal{P}\left(  \xi\right)  \frac{\partial\Theta}{\partial\xi}=\left(
\mathbf{\hat{P}}_{\bot}^{2}+m^{2}\right)  \Theta\,.\label{cf.18}%
\end{equation}
We can see that the spinor $\Theta\left(  \xi,x^{1},x^{2}\right)  $ may be
chosen as%
\begin{equation}
\Theta\left(  \xi,x^{1},x^{2}\right)  =\Phi_{\lambda}\left(  \xi,x^{1}%
,x^{2}\right)  \vartheta\,,\label{cf.19}%
\end{equation}
where the function $\Phi_{\lambda}\left(  \xi,x^{1},x^{2}\right)  $ obeys
equation (\ref{cf.20}) and $\vartheta$ is an arbitrary constant two-component
spinor. Then the transformation (\ref{cf.21}) allows one to reduce the problem
to the two-dimensional Schr\"{o}dinger equation (\ref{cf.14}) for a charged
particle in a constant uniform magnetic field. Solutions of the latter
equation are studied in detail: see, for example, \cite{15}.

\begin{acknowledgement}
V.B. thanks FAPESP for support and Nuclear Physics Department of S\~{a}o Paulo
University for hospitality; he also thanks Russia President grant
SS-1743.2003.2 and RFBR grant 03-02-17615 for partial support. D.G. is
grateful to FAPESP and CNPq for permanent support.
\end{acknowledgement}

\end{document}